\documentclass[12pt]{iopart}

\usepackage{amssymb}
\usepackage{color}
\usepackage[dvips]{graphicx}
\usepackage[margin=20pt,font=small,labelfont=bf]{caption}

\newcommand{\bq}{\begin{equation}}
\newcommand{\eq}{\end{equation}}
\newcommand{\ba}{\begin{eqnarray}}
\newcommand{\ea}{\end{eqnarray}}

\newcommand{\unity}{1 \hskip-2.5pt{\rm l}}
\newcommand{\up}{\uparrow}
\newcommand{\down}{\downarrow}

\def\bra#1{\ensuremath{\langle{#1}\vert}}
\def\ket#1{\ensuremath{\vert{#1}\rangle}}
\def\Landau{\mathcal{O}}
\def\kB{\ensuremath{k_{\rm B}}}
\def\expect#1{\ensuremath{\langle{#1}\rangle}}

\begin{document}

\parindent0pt

\title{Efficient creation of multipartite entanglement in flux qubits}

\author{J.~Ferber$^{1,2}$ and F.K.~Wilhelm $^{2,3,4}$}
\address{$^1$ Institut f\"ur Theoretische Physik,
Johann Wolfgang Goethe-Universit\"at,
Max-von-Laue-Str. 1,
60438 Frankfurt am Main, Germany}
\address{$^2$ Arnold Sommerfeld Center and Department Physik,  Ludwig-Maximilians-University, Theresienstr.\ 37, 80333 Munich, Germany.}
\address{$^3$ Institute for Quantum Computing and Department of Physics and
Astronomy, University of Waterloo, 200 University Avenue West,
Waterloo, ON, Canada, N2L 3G1}
\address{$^4$ Kavli institute for theoretical physics, University of California Santa Barbara, USA}
\ead{fwilhelm@iqc.ca}

\begin{abstract}
We investigate three superconducting flux qubits coupled in a loop. In this setup, tripartite entanglement can be created
in a natural, controllable, and stable way. Both generic kinds of tripartite entanglement --the W type as well as the GHZ type entanglement-- can be identified among the eigenstates. We also discuss the violation of Bell inequalities in this system and show the impact of a limited measurement fidelity on the detection of entanglement and quantum nonlocality.
\end{abstract}


\maketitle

\section{Introduction}

Entanglement is one of the most intriguing consequences of quantum mechanics \cite{Peres93} and has long been debated in a  case against its completeness \cite{EPR35}. Entanglement manifests the nonlocality of quantum mechanics. Bipartite entanglement is well established in optical systems, see, e.g., \cite{Weihs98,Kim09,Sackett00},  and has recently been verified in superconducting circuits \cite{Ansmann09,Chow09}. Quantum information processing uses entanglement as a resource \cite{Nielsen00,Horodecki09}, i.e., the preparation of entangled states is essential for quantum computing. 

In multipartite systems, different types of entanglement can be classified.  Specifically, in a tripartite system, one can recover the same type of bipartite entanglement as in a two-particle system -- the measurement of one of the particles reduces entanglement by taking out the particle being measured but leaves the entanglement between the remaining particles intact. On the contrary, there can also be tripartite entanglement where a single one-particle measurement completely destroys entanglement between all parties. 

Quantum computing and control is very mature in liquid-state nuclear magnetic resonance and in atomic and optical physics. In the former, due to the use of pseudopure states, it is not obvious that strong entanglement can be created. Creation of entangled states of ions, neutral atoms, and photons \cite{Weihs98,Kim09,Sackett00,Ansmann09,Chow09} has been highly successful. The required experiments, though, become increasingly complex, caused by the very weak interaction of the objects to be entangled, which often is also only effective for a very short time. 

On the other side, condensed matter systems typically have strong inter-particle interactions which can act effectively because the particles occupy fixed positions in space,  e.g. in a lattice. Thus, entangled states are not that exotic in correlated condensed systems \cite{Verstraete04}. Going beyond the simplest system, namely the spin singlet, we consider a single triangle of spins with antiferromagnetic Ising coupling. Without external field, the ground state of the system is spanned by the three degenerate frustrated states in which the orientation of one of the spins differs from the other two. Creating a superposition of these states right away leads to maximally entangled Werner (W) states \cite{Werner89} which only contain bipartite entanglement. These superpositions can be created in the quantum version of this model, i.e., if the system is put into a transversal magnetic field. Spin Hamiltonians with 'designed' properties can be implemented using macroscopic devices such as superconducting qubits rather than elementary spins \cite{Shumeiko06,Insight}. In particular, superconducting flux qubits allow for a strong inductive qubit-qubit interaction. 

The paper is organized as follows: In section \ref{section:flux_qubit_design} we propose a design for the coupling of three flux qubits. It provides a strong interaction whose strength can be designed over a large range during fabrication. The form of the eigenstates is discussed in section \ref{section:eigenstates}, states in the proximity of Greenberger-Horne-Zeilinger (GHZ) states \cite{GHZ1989} as well as W states can be identified. Section \ref{section:entanglement_features} describes the characterization and detection of tripartite entanglement in this system. In section \ref{section:Bell_inequalities}, we discuss the violation of Bell inequalities by virtue of optimally chosen operators and compare the required measurement fidelities with a representative two-qubit case.

\section{Flux qubit design}
\label{section:flux_qubit_design}

\begin{figure}[htb]
\centering
\includegraphics[width=0.6\columnwidth]{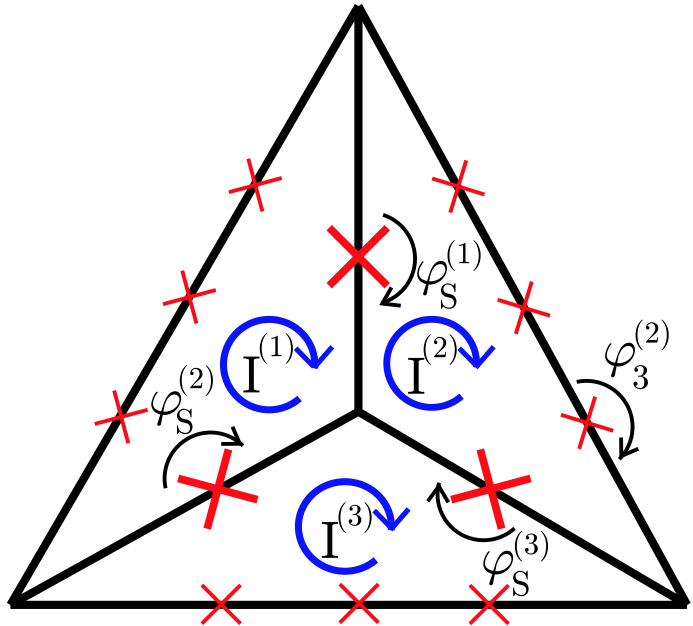}
\caption{The design of the
flux qubit triangle. The three qubits are formed by the three
small isosceles triangles, the round arrows in the qubits define
the directions of the currents. Small crosses represent the
Josephson junctions in the individual qubits, large crosses the
coupling junctions.}
\label{fig:triangle}
\end{figure}

We consider the triangle design sketched in figure \ref{fig:triangle}, consisting of three three-Josephson-junction (3JJ) flux qubits \cite{Mooij1999}. The individual qubits operate in the flux regime with $\kB T \ll E_{\rm C}^{(i)} \ll E_{\rm J}^{(i)} \ll \Delta$, $i=1,\:2,\:3$, and $\Delta$, $E_{\rm C}^{(i)}$, and $E_{\rm J}^{(i)}$ are the superconducting gap, the charging and Josephson energy of the $i$-th qubit, respectively. Therefore, the two different directions of persistent currents in the qubit loop form the basis states of the qubits. As this qubit design shows quantum behaviour even for small self-inductance, it allows for the fabrication of circuits that are relatively insensitive to external noise. However, a small self-inductance results in a very weak inductive coupling. This makes simple coupling schemes, like placing disconnected qubits close to each other, insufficient as they are restricted by the mutual geometric inductances. In contrast, the coupling between the qubits in this design is provided by shared lines with additional large Josephson junctions \cite{Plantenberg07}.

\medskip

As the potential energy is dominated by the Josephson energies of the junctions, we will only take this energy into account for the calculation of the coupling strength and neglect the inductive energy from the geometrical inductance of the triangle and the kinetic inductance of the shared lines. For convenience, we will consider the Josephson energy of the qubit junctions (the ones on the outer edges of the triangle) and the large shared junctions separately.

The total Josephson energy of the qubit junctions is given by
\bq
E_{\rm Jos,Q}=-E_{\rm J}\:\sum_{i=1}^3(\cos{\varphi_1^{(i)}}+\cos{\varphi_2^{(i)}}+\alpha\cos{\varphi_3^{(i)}})
\label{jos_energy_qubit_junctions}\;.\eq
Combining the additional phases across the shared junctions into the phase $\phi^{(i)}$ (e.g. $\phi^{(1)}=\varphi_S^{(1)}-\varphi_S^{(2)}$ for qubit 1), the fluxoid quantization reads
\bq \varphi_1^{(i)}+\varphi_2^{(i)}+\varphi_3^{(i)}+\phi^{(i)}+\frac{2 \pi
\Phi_{\rm x}^{(i)}}{\Phi_0}=0\;, \label{fluxoid_triangle}\eq
where $\Phi_{\rm x}$ denotes the externally applied flux.

The coupling junctions are large compared to the qubit junctions
and their critical currents are far above the persistent currents $I_{\rm p}$
in the qubits. Hence, their phases are small and we obtain for first order (for qubit 1)
\begin{equation}
\phi^{(1)}\approx \frac{2
I_{\rm p}^{(1)}-I_{\rm p}^{(2)}-I_{\rm p}^{(3)}}{I_{\rm C,S}}\;,
\end{equation}
where the critical current $I_{\rm C,S}$ is assumed to be equal for all three shared junctions.

In order to separate the effect of $\phi^{(i)}$, we solve the fluxoid quantization for, e.g., $\varphi_3^{(i)}$,
\bq \alpha \cos{\varphi_3^{(i)}}=\alpha \cos{\left(\frac{2 \pi
\Phi_{\rm x}^{(i)}}{\Phi_0}+\varphi_1^{(i)}+\varphi_2^{(i)}\right)}+\frac{I_{\rm p}^{(i)}}{I_{\rm C}}\phi^{(i)}\;,
\eq
where we again expanded to first order in $\phi^{(i)}$ and used $2\pi\Phi_{\rm x}^{(i)}/\Phi_0\approx \pi$ as well as the relation $\varphi_1^{(i)}=\varphi_2^{(i)}=\pm\varphi^*$ with $\cos{\varphi^*}=1/2\alpha$ \cite{Orlando1999} for the minima of the potential landscape of a single qubit.
Equation (\ref{jos_energy_qubit_junctions}) then can be rewritten as
\bq E_{\rm Jos,Q}=E_{\rm
Jos,uncp}-\frac{\hbar^2}{2e^2 E_{\rm J,S}}\sum_{i=1}^3 {I_{\rm p}^{(i)}}^2+
\frac{\hbar^2}{2e^2 E_{\rm J,S}} \sum_{i=1}^3 \sum_{j>i}I_{\rm p}^{(i)}I_{\rm p}^{(i)}\;,\eq
with the Josephson energy of the uncoupled qubit system $E_{\rm
Jos,uncp}$. This contribution to the coupling is thus found to be pairwise antiferromagnetic, with a larger shared junction resulting in a weaker coupling.

The second major contribution to the potential energy is the Josephson energy of the shared junctions \footnote{Note that we expand to second order in $\varphi_{\rm S}^{(i)}$ here; this is justified by $E_{\rm J,S}/E_{\rm J}=I_{\rm C,S}/I_{\rm C}$ which makes (\ref{jos_energy_shared_junctions}) and (\ref{jos_energy_qubit_junctions}) of the same order in $I_{\rm C}/I_{\rm C,S}$.},

\bq E_{\rm Jos,S}\!=\!-E_{\rm J,S}\sum_{i=1}^3\cos{\varphi_{\rm S}^{(i)}}\!\approx\!
-E_{\rm J,S}\sum_{i=1}^3\left(\!1-\frac{{\varphi_{\rm S}^{(i)}}^2}{2}\right)\label{jos_energy_shared_junctions}.\eq

Using $\varphi_{\rm S}^{(1)}\approx (I_{\rm p}^{(1)}-I_{\rm p}^{(2)})/I_{\rm C,S}$ etc., it reads
\bq E_{\rm Jos,S}=-3E_{\rm J,S}+\frac{\hbar^2}{4e^2 E_{\rm J,S}}\sum_{i=1}^3
{I_{\rm p}^{(i)}}^2-
\frac{\hbar^2}{4e^2 E_{\rm J,S}} \sum_{i=1}^3 \sum_{j>i}I_{\rm p}^{(i)}I_{\rm p}^{(j)} \;, \eq
yielding a ferromagnetic coupling with half the strength of the qubit term.

Hence the total the coupling strength is given by
\bq
\Delta U^{(ij)} \!\equiv\! C^{(ij)} \!=\! \frac{\hbar^2}{4e^2 E_{\rm J,S}} I_{\rm p}^{(i)}I_{\rm p}^{(j)}+\Landau[(E_{\rm J}/E_{\rm J,S})^2].
\eq
Expressing the persistent currents in the qubits in terms of the Josephson energy of the qubit junctions, $I_{\rm p}=I_{\rm C}\sqrt{1-1/4\alpha^2}=2e E_J/\hbar\cdot \sqrt{1-1/4\alpha^2}$, the coupling is seen to depend directly on the size ratio $r=E_{\rm J}/E_{\rm J,S}$ of the qubit's and shared junctions,
\bq
C=r E_{\rm J} \left(1-\frac{1}{4\alpha^2}\right)\label{coupling}.
\eq
In experiments with inductive coupling \cite{Plantenberg07} or shared lines \cite{Majer2005} between the qubits, the coupling is very small ($C/E_{\rm J}\approx 0.1\%-0.5\%$), reflecting the small mutual inductance (geometric and/or bulk superconducting kinetic inductance) that can be achieved in such designs. Contrastly, (\ref{coupling}) allows for coupling strengths which are one order of magnitude stronger. The shared junctions are still large and linear: for a size ratio $r$ of e.g. $1\%-5\%$, the shared junctions basically behave like a linear inductor,  $L_{\rm J,S}=\Phi_0/(2\pi I_{\rm C,S}\cos \phi_{\rm S})\approx\hbar^2/(4e^2 E_{J,S})$. Thus, the shared junctions can be viewed as ultra-compact inductors making the qubit loops equivalent to the standard 3-junction flux qubit with a moderate inductance \cite{Mooij1999,Robertson2006}.

\smallskip
The currents in the qubits are quantum mechanically associated with $\hat\sigma_z$ operators and the effective Hamiltonian reads in terms of the Pauli spin matrices

\bq
\mathbf{H}=\sum_{i=1}^3
\left(\!-\frac{1}{2}\epsilon_i\,\hat\sigma_z^{(i)}-\frac{1}{2}\Delta_i\,\hat\sigma_x^{(i)}
\right) + \sum_{i=1}^3 \sum_{j>i} C\hat\sigma_z^{(i)}\hat\sigma_z^{(j)}.
\label{eq:Hamiltonian}
\eq

The energy bias $\epsilon$ can be tuned by the externally applied magnetic flux, whereas the tunnel matrix element $\Delta$ depends on the size ratio of the qubits' junctions \cite{Orlando1999, Makhlin2001}. In the following we assume the qubits to be identical ($\Delta_1=\Delta_2=\Delta_3\equiv\Delta$, $\epsilon_1=\epsilon_2=\epsilon_3\equiv\epsilon$). Note that even though the tunnel splittings $\Delta_i$ exponentially depend on fabrication parameters
\cite{Orlando1999}, recent experiments \cite{Hime06} have made them extremely close to each other and, in principle, their values can be externally adjusted \cite{Paauw09}.

\section{Eigenstates of the system}
\label{section:eigenstates}

\begin{figure}[htb]
\centering
\includegraphics[width=0.9\columnwidth]{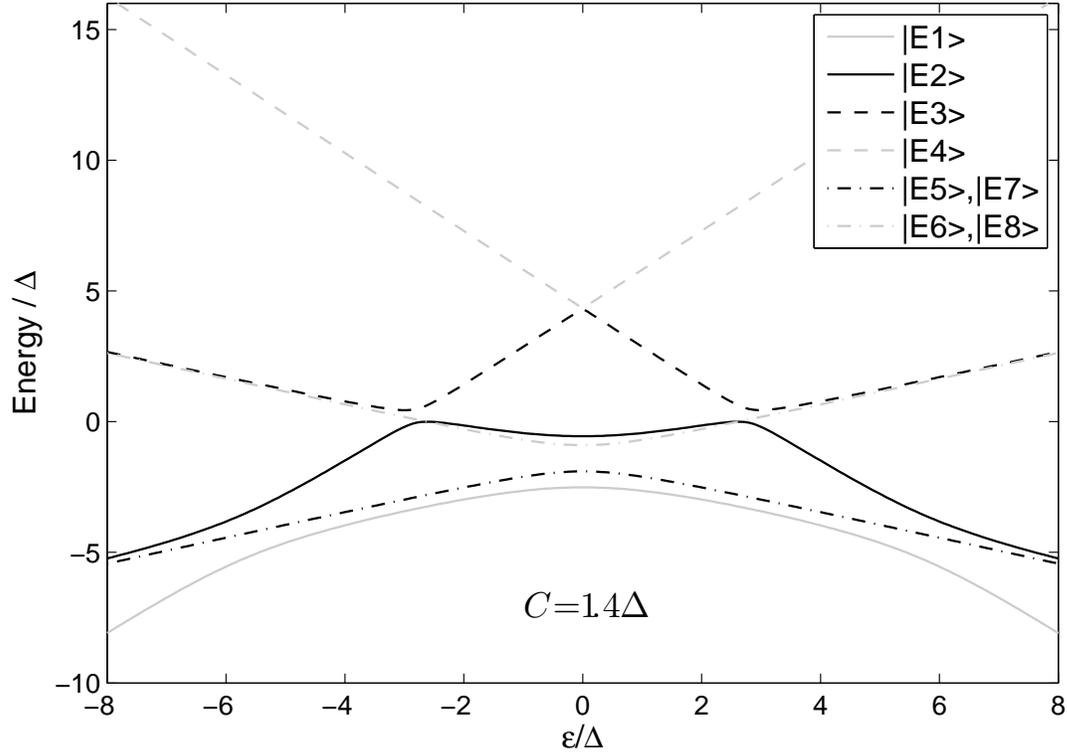}
\caption{Plot of the eigenenergies of the eigenstates
$\ket{E_1}$--$\ket{E_8}$ for the indicated coupling strength. $\ket{E_5}$ and $\ket{E_6}$ as well as $\ket{E_7}$ and $\ket{E_8}$ are always degenerate.}
\label{fig:energy_level_diagram}
\end{figure}

We aim for preparing tripartite entangled states in a preferably easy and stable way. Both
demands are naturally met by the eigenstates of a system, as they are easy to prepare by
$\pi$-pulse driving and stable against pure dephasing processes. Figure \ref{fig:energy_level_diagram} displays the eigenenergies as a function of the energy bias $\epsilon$ for a given coupling strength. These have been found numerically but can be understood analytically. 

For large positive or negative energy bias, the ground state and the highest excited state are the classical, totally polarized states $\ket{\up\up\up}$ or $\ket{\down\down\down}$, respectively, whatever is favored by the magnetic field. In this case, the antiferromagnetic coupling can be disregarded. The other excited states are accordingly the classical frustrated states with the three states $\ket{\up\down\down}$, $\ket{\down\up\down}$, and $\ket{\down\down\up}$ being degenerate, as well as $\ket{\down\up\up}$, $\ket{\up\down\up}$, and $\ket{\up\up\down}$.

\smallskip

Going to finite energy bias, we need to take the coupling into account ($C=1.4\,\Delta$ in the following). The antiferromagnetic coupling energetically favors frustrated states. For $\epsilon=0$, the ground state $\ket{E_1}$ therefore contains a larger contribution of frustrated states and a smaller contribution of polarized states,

\bq
\ket{E_1}= \frac{1}{\sqrt{6+2\,\delta^2}}\big\{\ket{f_{\rm all}} +\delta(\ket{\up\up\up}+\ket{\down\down\down})\big\}\;,
\label{eq:groundstate_zero_energy_bias}
\eq

where $\ket{f_{\rm all}}$ denotes the equal (non-normalized) superposition of all frustrated states

\bq
\ket{f_{\rm all}}=\ket{\up
\down\down}+\ket{\up\down\up}+\ket{\up\up\down}+\ket{\down\up\up}+\ket{\down\up\down}+\ket{\down\down\up}
\;,
\eq

and $\delta\approx0.2$ is small, i.e. the aligned states $\ket{\up\up\up}$ and $\ket{\down\down\down}$ are suppressed.

The highest excited states $\ket{E_3}$ and $\ket{E_4}$ show the opposite behaviour and consist for $\epsilon=0$ mainly of a superposition of distinct polarized states,

\bq
\ket{E_3}=\frac{1}{\sqrt{2+6\,\delta_1^{\:2}}}\big\{\ket{\down\down\down}+\ket{\up\up\up}-\delta_1 \ket{f_{\rm all}}\big\}
\eq

with $\delta_1\approx 0.07$, and

\bq
\ket{E_4}=\frac{1}{\sqrt{2+6\,\delta_2^{\:2}}}\big\{\ket{\down\down\down}-\ket{\up\up\up}+\delta_2\ket{f_{\rm all}}\big\}
\eq

with $\delta_2\approx 0.1$. Thus, we find these states to be in the proximity of GHZ states, commonly represented by $\ket{\rm GHZ}=(\ket{\down\down\down}\pm\ket{\up\up\up})/\sqrt{2}$.

Due to the large antiferromagnetic coupling, the eigenstates discussed above evolve into classical states already for small detuning of the energy bias. However, there are more regimes of tripartite entangled states among the spectrum of eigenstates: at finite positive and negative energy bias $\epsilon=\pm\epsilon\ast\approx \pm 2.6 \Delta$ two more anticrossings involving $\ket{E_2}$ and $\ket{E_3}$ are present. The explicit form of the state forming the lower branch at $\epsilon=-\epsilon\ast$ reads
\ba
\ket{E_2}_{-\epsilon*} &=&\frac{1}{2\sqrt{1+\delta_3^{\:2}}}\big\{\ket{\up\up\down}+\ket{\up\down\up}+\ket{\down\up\up}-\ket{\down\down\down} \big\}+\nonumber\\
&&+\delta_3\big\{\ket{\down\down\up}+\ket{\down\up\down}+\ket{\up\down\down}-\ket{\up\up\up} \big\}\:,
\label{eq:e2_at_left_maximum}
\ea

where $\delta_3\approx 0.09$. Therefore, this state is close to

\bq
\ket{\overline{\rm GHZ}}=\frac{1}{2} \big(\ket{\up\up\down}+\ket{\up\down\up}+\ket{\down\up\up}-\ket{\down\down\down}\big)=
\frac{1}{\sqrt{2}}\big(\ket{\bar{0}\bar{0}\bar{0}}+\ket{\bar{1}\bar{1}\bar{1}}\big)
\label{eq:ghzbar}
\eq

with $\ket{\bar{0}}=(\ket{\up}+i\ket{\down})/\sqrt{2}$ and $\ket{\bar{1}}=-(\ket{\up}-i\ket{\down})/\sqrt{2}$. As $\ket{\overline{\rm GHZ}}$ and $\ket{\rm GHZ}$ can be transferred onto each other by purely local operations, they have identical entanglement properties.

\medskip

The dash-dot lines in figure \ref{fig:energy_level_diagram} indicate two two-fold degenerate subspaces. It is shown in the appendix that arbitrary states in these two-dimensional Hilbert spaces can be prepared by coupling these subspaces to the ground state via resonant driving. The subspaces are spanned by frustrated states. Since any frustrated state $\ket{f}$ is an eigenstate of the coupling operator,
$( \hat\sigma_z^{(1)} \hat\sigma_z^{(2)} +
\hat\sigma_z^{(1)} \hat\sigma_z^{(3)} + \hat\sigma_z^{(2)}
\hat\sigma_z^{(3)} ) \ket{f}=-\ket{f}$,
the form of an eigenstate prepared in these subspaces does not change with the coupling strength. Among the eigenstates, W type states can be found, with the maximally entangled state (with respect to the measure of \emph{global entanglement} \cite{Endrejat2005}) in the lower energy subspace for $\epsilon=0$ taking the form
\bq\fl
\ket{\psi_{\rm max}^{\rm L}}\!=\!\frac{1}{2\sqrt{6}}\Big\{ 2(\ket{\up\up\down}+\ket{\down\down\up})-
(1-i\sqrt{3})(\ket{\up\down\up}+\ket{\down\up\down})-
(1+i\sqrt{3})(\ket{\up\down\down}+\ket{\down\up\up})\Big\}.
\eq

The local transformation $\hat U^{\rm L}\,\ket{\psi_{\rm max}^L}=\ket{W}$ rotating $\ket{\psi_{\rm max}^{\rm L}}$ onto the common representation of a W state, $\ket{W}=\frac{1}{\sqrt{3}}(\ket{\up\up\down}+\ket{\up\down\up}+\ket{\down\up\up})$, is given by
\bq\fl
\hat U^{\rm L}=\{\rme^{-\rmi \pi/3}\,\hat R_y(\pi/2)\}^{(1)}
\otimes\{\hat R_z(2\pi/3)\:\hat R_y(\pi/2)\}^{(2)}
\otimes\{\hat R_z(-2\pi/3)\:\hat R_y(\pi/2)\}^{(3)}\;,
\label{eq:propagator_U_L}
\eq

where $\hat R_z(\theta)$ ($\hat R_y(\theta)$) is a rotation around the z-axis (y-axis) by
an angle $\theta$. Thus, we can see explicitly that the state $\ket{\psi_{\rm max}^L}$ is indeed locally equivalent to a W state.

With respect to the occurrence of W states, the relative pairwise coupling strengths between the qubits are essential: equal mutual (antiferromagnetic) couplings like in the design proposed here make the system frustrated and facilitate W-like eigenstates. This is not given in, e.g, a linear chain of qubits where the two nearest-neighbour couplings are stronger than the next-nearest neighbour coupling between the outer qubits.

\section{Detection of tripartite entanglement}
\label{section:entanglement_features}

Conjecturing entanglement from the energy spectrum is not a satisfactory experimental 
indication. We thus want to discuss measurement protocols to verify entanglement more explicitly. 

A necessary and sufficient condition for the entanglement of a given state is only known for two-qubit systems \cite{Peres1996}. The determination of entanglement for multipartite states, however, is an open question. For three qubits two inequivalent kinds  \cite{Cirac2000} of \emph{genuine} multipartite entanglement (i.e. each party is entangled with each other party) can occur, represented by the GHZ state and the W state. A tool for detection of any kind of genuine tripartite entanglement for arbitrary states is not at hand; however, if some knowledge about the state under investigation is provided, entanglement witnesses (EWs) can be used \cite{Terhal2000, Lewenstein2000}. These are observables with a positive expectation value for all (bi-)separable states (in general $n-1$ partite entangled states), whereas a negative expectation value indicates the presence of tripartite ($n$-partite) entanglement. The common way to construct an EW $\mathcal{W}$ for a state $\ket{\psi}$ is
\bq
\mathcal{W}=\alpha\,\unity-\ket{\psi}\bra{\psi}\;,
\label{entanglement_witnesses_construction}
\eq

where $\alpha$ is the maximal squared overlap of $\ket{\psi}$ with any biseparable or fully separable state. Determination of $\alpha$ is in general complicated \cite{Lewenstein2000, Eisert2004, Brandao2005}, but we can use the aforementioned proximity of the states under investigation to W and GHZ states, respectively, to make use of known values for $\alpha$. We will discuss how to measure EWs in the end of this section.

\begin{figure}[htb]
\centering
\includegraphics[width=0.8\columnwidth]{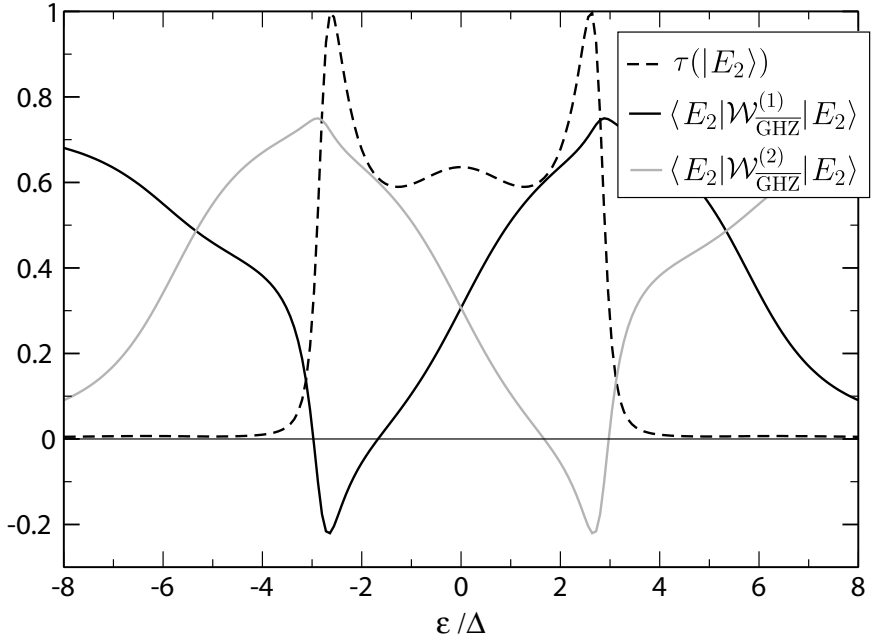}
\caption{3-tangle and expectation value of the GHZ witnesses $\mathcal{W}_{\overline{\rm GHZ}}^{(1)}$ and $\mathcal{W}_{\overline{\rm GHZ}}^{(2)}$ for the state $\ket{E_2}$. For finite energy bias $\epsilon=\pm \epsilon*$ we find a peaking 3-tangle as well as negative expectation value for the two GHZ witnesses, indicating entanglement of the GHZ type. Moreover, the entanglement is relatively robust to detuning of the energy bias.}
\label{fig:e2_3_tangle_ghz_witness}
\end{figure}

Figure \ref{fig:e2_3_tangle_ghz_witness} shows the expectation values for two EWs $\bra{E_2}\mathcal{W}_{\overline{\rm GHZ}}^{(1)}\ket{E_2}$ and $\bra{E_2}\mathcal{W}_{\overline{\rm GHZ}}^{(2)}\ket{E_2}$ for varying energy bias. The EW $\mathcal{W}_{\overline{\rm GHZ}}^{(1)}$ is constructed such as to detect states of the form given in (\ref{eq:e2_at_left_maximum}), i.e., $\ket{E_2}$ at the left anticrossing. In order to construct the optimal EW, one would have to determine the maximal squared overlap of $\ket{E_2}$ in this point with any non-GHZ entangled state. Instead, we make use of the vicinity of $\ket{E_2}_{-\epsilon*}$ with $\ket{\overline{\rm GHZ}}$  defined in (\ref{eq:ghzbar}) and use \cite{Acin2001} $\mathcal{W}_{\overline{\rm GHZ}}^{(1)}=\frac{3}{4} \,\unity-\ket{\overline{\rm GHZ}}\bra{\overline{\rm GHZ}}$. The optimal local decomposition of $\mathcal{W}_{\overline{\rm GHZ}}^{(1)}$ requires four experimental settings \cite{Weinfurter2004}, see \ref{ch:appendix_local_measurements}. Since the system is invariant under a combined flip of the spins and an inversion of the sign of $\epsilon$, $\mathcal{W}_{\overline{\rm GHZ}}^{(2)}$ (the optimal EW at the right anticrossing) is the totally flipped counterpart to $\mathcal{W}_{\overline{\rm GHZ}}^{(1)}$, $\mathcal{W}_{\overline{\rm GHZ}}^{(2)}=\hat R_x(\pi)^{\otimes 3}\,\mathcal{W}_{\overline{\rm GHZ}}^{(1)}\,\hat R_x(-\pi)^{\otimes 3}$.

\smallskip

Figure \ref{fig:e2_3_tangle_ghz_witness} also shows the 3-tangle $\tau$ \cite{Wootters2000}, which allows for a reliable distinction between the two classes of entanglement, as it is zero for all W type states (and all separable states, of course), whereas it takes positive values for all states in the GHZ class. However, the 3-tangle can only be measured by full state tomography which requires many more settings than the witness.
Both quantities indicate a strong (limiting case: $\tau_{\rm max}=\tau(\ket{\rm GHZ})=1$, $\expect{\mathcal{W}_{\overline{\rm GHZ}}^{(1)}}_{\min} =\expect{\mathcal{W}_{\overline{\rm GHZ}}^{(2)}}_{\min}=-1/4$) tripartite entanglement of GHZ type over a large range of the energy bias.

\smallskip

\begin{figure}[htb]
\centering
\includegraphics[width=0.7\columnwidth]{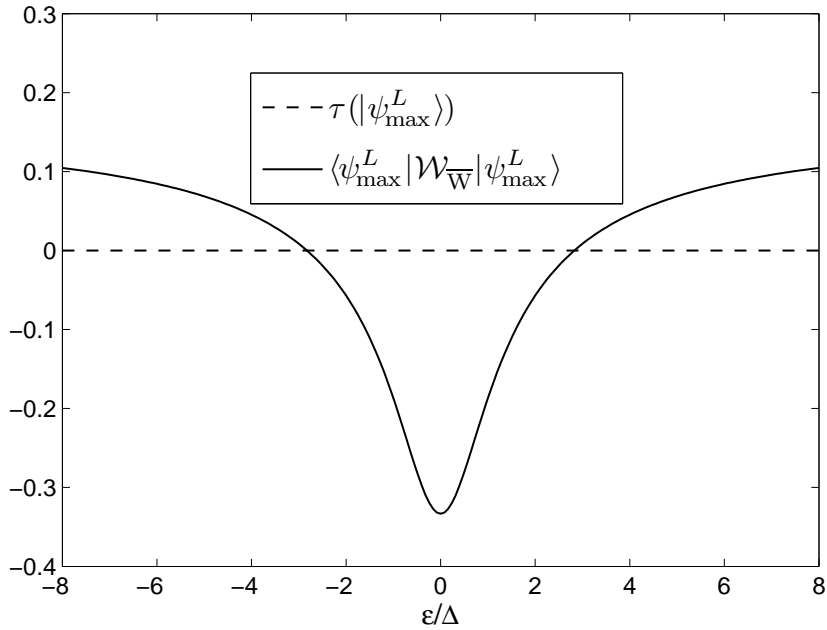}
\caption{3-tangle and expectation value of the W witness $\mathcal{W}_{\rm \overline{W}}$ for the state $\ket{\psi_{\rm max}^{\rm L}}$. The vanishing 3-tangle excludes entanglement of the GHZ type, whereas the negative expectation value of the W witness indicates a W type entanglement.}
\label{fig:e5_3_tangle_w_witness}
\end{figure}

In Figure \ref{fig:e5_3_tangle_w_witness}, the 3-tangle and the expectation value of the corresponding EW for $\ket{\psi_{\rm max}^{\rm L}}$, $\mathcal{W}_{\rm \overline{W}}=\frac{2}{3}\,\unity - \ket{\psi_{\rm max}^{\rm L}} \bra{\psi_{\rm max}^{\rm L}}$ \cite{Acin2001} is displayed. Its expectation value is positive for biseparable and fully separable states.

It thus detects genuine tripartite entanglement in general, without distinguishing between W and GHZ type entanglement. In connection with the 3-tangle a distinction can be achieved, though, stating an entanglement of the W type in a large range for $\epsilon$, approaching its theoretical minimum ($\expect{\mathcal{W}_{\rm \overline{W}}}=-1/3$) for zero energy bias.

Measuring these entanglement witness only requires single-qubit measurements of certain linear combinations of Pauli matrices. 
As a consequence, in a superconducting qubit setting, this can be achieved by decoupling the qubits, performing an appropriate single-qubit rotation, and then measuring the most accessible variable such as the flux, here corresponding to $\hat{\sigma}_z$. 
Finding such a local decomposition is a demanding task that has been solved in literature \cite{Guehne2003, Weinfurter2004}.
The appropriate optimal local decompositions for the witnesses employed here are shown in \ref{ch:appendix_local_measurements}. Local measurements of this type have been performed in the context of Bell tests \cite{Ansmann09,Chow09}.

We have to note that in the proposed setup, the couplings between qubits are permanently switched on. However, this coupling can be effectively removed by nonadiabatically pulsing $\epsilon$ to strongly detuned values. The evolution in this detuned state corresponds to single-qubit rotations alone and thus does not change entanglement.

\section{Violation of Bell inequalities and robustness to limited measurement fidelity}
\label{section:Bell_inequalities}

So far, we made use of EWs as tool for the detection of tripartite entanglement. Another common approach for the detection of entanglement are Bell inequalities. Multiqubit states can contradict local realistic models in a new and stronger way than two-qubit states \cite{GHZ1990, Mermin1990}, reflected by a stronger violation of Bell inequalities \cite{MerminPRL1990, Klyshko1993}. In the case of three qubits, quantum mechanics predicts $\bra{\rm GHZ}\hat M_{\rm GHZ}\ket{\rm GHZ}=4$ for the expectation value of a Bell operator $\hat M_{\rm GHZ}$ in the GHZ state \footnote{the specific form of $\hat{M}_{\rm GHZ}$ will be given later}, while the local prediction gives an upper threshold of $\bra{\psi}\hat M_{\rm GHZ} \ket{\psi}\le 2$ for any states $\ket{\psi}$. In comparison, the maximal expectation value of a Bell operator for a two-qubit state is $2\sqrt{2}$, in contrast to the local prediction of $\le 2$, yielding a maximal violation only by a factor of $\sqrt{2}$. However, Bell inequalities are not distinctive to the type of entanglement; moreover, there are entangled states that do not violate any Bell inequality. Nevertheless, their violation as sign for non-classical correlations is highly substantial as an ingredient to quantum information processing. Besides, it also allows for a comparison between the twopartite case and the tripartite case with respect to the robustness to limited measurement fidelity.

\smallskip

\begin{figure}[htb]
\centering
\includegraphics[width=0.7\columnwidth]{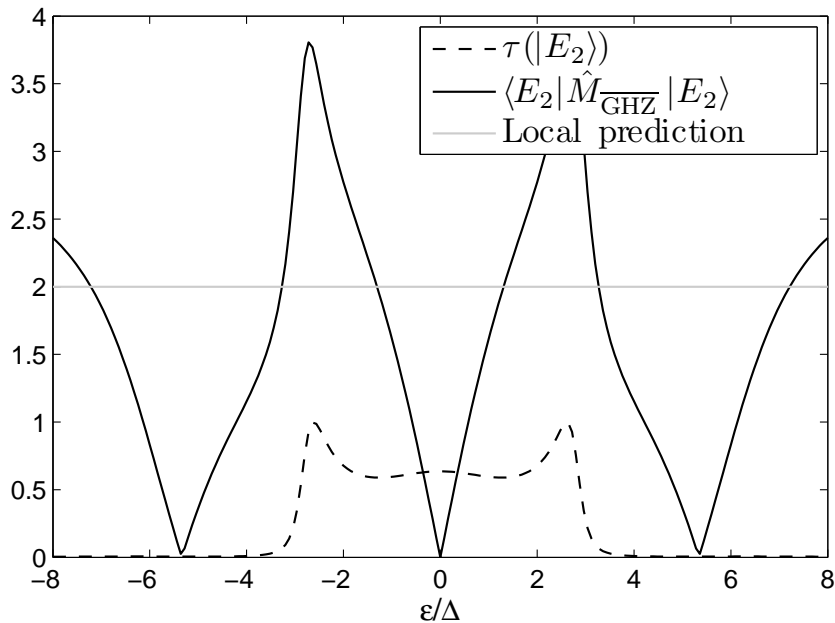}
\caption{3-tangle and expectation value of the Bell operator $\hat{M}_{\rm \overline{GHZ}}$ for the state $\ket{E_2}$. A significant violation of the corresponding Bell inequality can be observed.}
\label{fig:e2_3_tangle_bell_inequ}
\end{figure}

\begin{figure}[htb]
\centering
\includegraphics[width=0.6\columnwidth]{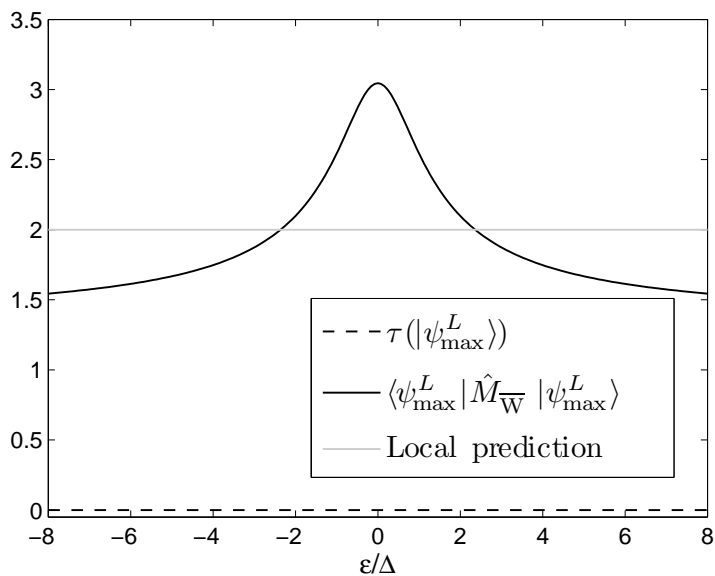}
\caption{3-tangle and expectation value of the Bell operator $\hat{M}_{\rm \overline{W}}$ for the state $\ket{\psi_{\rm max}^L}$. The maximal violation of the Bell inequality is not as high as for $\ket{E_2}$ in figure \ref{fig:e2_3_tangle_bell_inequ}, however, the violation persists over a larger range of $\epsilon$.}
\label{fig:e5_3_tangle_bell_inequ}
\end{figure}

We again investigate the two states $\ket{E_2}$ and $\ket{\psi_{\rm max}^{\rm L}}$ for varying energy bias, see figures \ref{fig:e2_3_tangle_bell_inequ} and \ref{fig:e5_3_tangle_bell_inequ}. For $\ket{E_2}$, we use the Bell operator \cite{MerminPRL1990} $\hat M_{\rm \overline{GHZ}}=
\hat\sigma_x\,\hat\sigma_x\,\hat\sigma_z+\hat\sigma_x\,\hat\sigma_z\,\hat\sigma_x+\hat\sigma_z\,\hat\sigma_x\,\hat\sigma_x-\hat\sigma_z\,\hat\sigma_z\,\hat\sigma_z$.
As for EWs, we can determine optimal Bell operators for given target states. In general, Bell inequalities for three qubits are constructed from the operator $\hat m(\bi{a},\bi{b},\bi{c})=(\bi{a}\cdot \hat{\bi{\sigma}})\otimes (\bi{b}\cdot \hat{\bi{\sigma}})\otimes(\bi{c}\cdot \hat{\bi{\sigma}})$, where $\bi{a}$, $\bi{b}$, and $\bi{c}$ are real three-dimensional normalized vectors, which define a rotation of the Pauli matrices $\hat{\bi{\sigma}}=(\hat\sigma_x, \hat\sigma_y, \hat\sigma_z)$. A Bell operator is then given by
\bq
\hat M=\hat m(\bi{a},\bi{b},\bi{c'})+\hat m(\bi{a},\bi{b'},\bi{c})+\hat m(\bi{a'},\bi{b},\bi{c})-\hat m(\bi{a'},\bi{b'},\bi{c'}).
\eq

In order to obtain an optimal Bell operator for a given state, one can optimize over the six unit vectors $\bi{a}$, $\bi{a'}$, $\bi{b}$, $\bi{b'}$, $\bi{c}$, and $\bi{c'}$ \cite{Endrejat2005, Emary2004}. The optimal values for $\hat M_{\rm W}$ adapted to the state $\ket{\rm W}$, as obtained by a numerical optimization, are listed in table \ref{tab_optimal_bell_params}.
\begin{table}[htb]
\caption{\label{tab_optimal_bell_params} Components of the vectors $\bi{a},\,\bi{a'},\,\bi{b},\,\bi{b'},\,\bi{c}$ and $\bi{c'}$ for the Bell operator $\hat M_{\rm W}$.}
\centering
\begin{tabular}{lrlrlr}
\br
$a_1$ & 0.318 & $b_1$ & 0.635 & $c_1$ & 0.635\\ 
$a_2$ & 0.250 & $b_2$ & 0.501 & $c_2$ & 0.501\\
$a_3$ & 0.914 & $b_3$ & -0.587 & $c_3$ & -0.587\\
\mr
$a'_1$ & -0.635 & $b'_1$ & 0.318 & $c'_1$ & 0.318\\
$a'_2$ & -0.501 & $b'_2$ & 0.250 & $c'_2$ & 0.250\\
$a'_3$ & -0.587 & $b'_3$ & 0.914 & $c'_3$ & 0.914\\
\br
\end{tabular}
\end{table} 
The Bell operator for the state $\ket{\psi_{\rm max}^{\rm L}}$ then reads $\hat M_{\rm \overline{W}}=\hat U^{\rm L \, \dag}\, \hat{M}_{\rm W}\, \hat U^{\rm L}$ with $U^{\rm L}$ being the local propagator from (\ref{eq:propagator_U_L}).

\medskip

Any experimental test of tripartite entanglement or the violation of Bell inequalities involving three qubits will be more fragile than a two particle test and will be put in jeopardy by detector imperfections (as three-party correlations need to be measured in either case, the measurement fidelity enters with the third power) and fabrication uncertainties. However, for the Bell inequalities, the stronger violation might compensate for that. We consider the effect of a limited measurement fidelity $f<1$ on the expectation values of the EWs and Bell operators introduced above and compare the results to a representative two-particle case. We model a non-perfect measurement of a spin component $\hat\sigma_i$ by the perfect measurement of a spin component $\hat\sigma'_i$ which yields the correct measurement result with a probability $f$ and '1' otherwise,
\bq
\hat\sigma'_i=f\hat\sigma_i+(1-f)\unity\;.
\eq

\begin{figure}[htb]
\centering
\includegraphics[width=0.8\columnwidth]{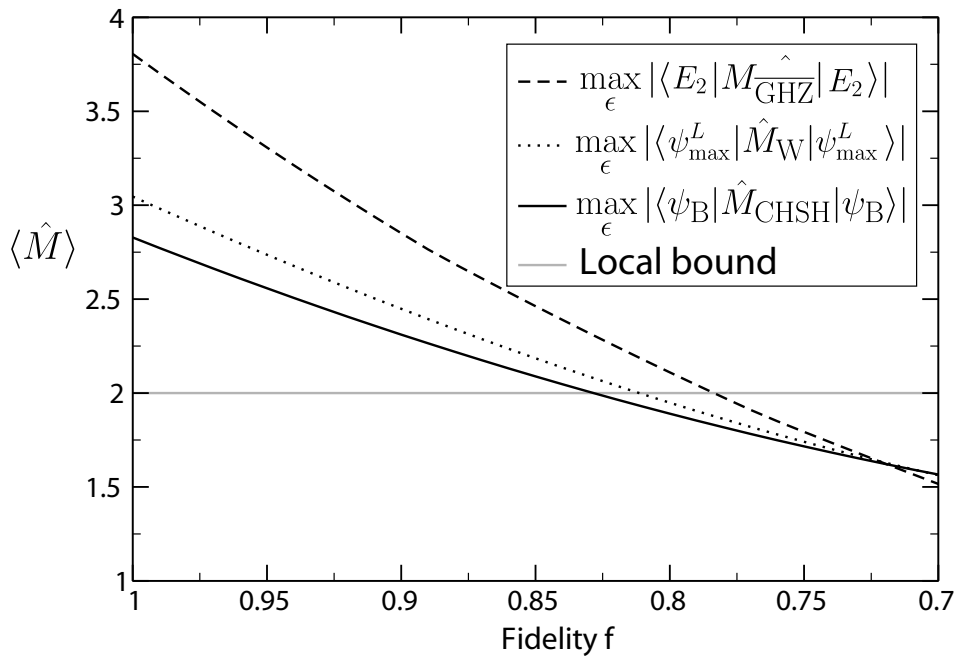}
\caption{Expectation values of optimal Bell operators for GHZ type, W type, and CHSH type entanglement shown as a function of decreasing measurement fidelity.}
\label{fig:max_violation_vs_fidelity_2_5_CHSH}
\end{figure}

Figure \ref{fig:max_violation_vs_fidelity_2_5_CHSH} compares the decay of the Bell violation with decreasing measurement fidelity for the cases discussed above and for a representative two-qubit case, namely the violation of the Clauser-Horne-Shimony-Holt (CHSH) inequality \cite{CHSH1969}) by a Bell pair $\ket{\psi_{\rm B }}=\frac{1}{\sqrt{2}}(\ket{\up\down}-\ket{\down\up})$. We find that the larger initial violation for the tripartite cases compensates for the quicker decay of the violation and allows for a slightly lower minimal detector fidelity; table \ref{tab:min_fidelities} summarizes the minimal detector fidelities for the detection of tripartite entanglement or violation of Bell inequalities. The requested measurement fidelity is already available for charge qubits, where significant progress has been achieved with dispersive readout inside a cavity, providing a visibility of more than 90$\%$ \cite{Wallraff2005}. A similar design has been proposed for flux qubits \cite{Markus_phd}. Moreover, other experiments based on Josephson junction technology indicating similar fidelities have been performed \cite{Lupascu2004, Lupascu2005, Devoret2004, Astafiev2004,Ansmann09}.

\begin{table}[htb]
\caption{\label{tab:min_fidelities}Minimal detector fidelities for the detection of tripartite entanglement or violation of Bell inequalities, respectively.}
\centering
\begin{tabular}{llll}
\br
  Operator & $f_{\rm min}$ & Operator & $f_{\rm min}$\\
\mr
$\mathcal{W}_{\rm GHZ}$ & $84.3\%$ & $\hat M_{\rm GHZ}$ & $81.4\%$\\
$\mathcal{W}_{\overline{\rm GHZ}}^{(2)}$ & $88.2\%$ & $\hat M_{\rm \overline{GHZ}}$ & $78.4\%$\\
$\mathcal{W}_{\rm \overline{W}}$ & $86.1\%$ & $\hat M_{\rm \overline{W}}$ & $81.2\%$\\
& & $\hat M_{\rm CHSH}$ & $\frac{2}{1+\sqrt{2}}\approx 82.8\%$\\
\br
\end{tabular}
\end{table} 

\section{Related work}

Related work has shown similar properties for a ring of exchange-coupled qubits \cite{Roethlisberger08} even in the ground state. Open/linear coupling topologies, albeit easier to prepare experimentally, require more complex pulse sequences \cite{Galiautdinov08,Wei06,Matuso07}  because the eigenstates do not have tripartite entanglement; they become more efficient in connected networks \cite{Galiautdinov09}. Also, tripartite entanglement between two superconducting cavities and one qubit has been proposed \cite{Switch}. Beyond tripartite entanglement, a circuit QED setup has been suggested for the fast preparation of an $N$-qubit GHZ state in superconducting flux or charge qubits \cite{Wang2009}.

\section{Conclusions}

In this paper, we introduced a system of three coupled flux qubits in a loop. We showed that it exhibits
strong tripartite entanglement for a realistic and approachable set of parameters
and that it is possible to detect and quantify this entanglement.

We acknowldege support by the Deutsche Forschungsgemeinschaft through SFB 631 and the NSERC disovery grants program and, in parts, by the National Science Foundation under Grant No. NSF PHY05-51164.

\appendix

\section{State preparation in a degenerate subspace \label{ch:appendix}}

We describe here, how the highly entangled states (e.g., $\ket{\psi_{\rm max}^L}$) in the degenerate subspaces introduced in section \ref{section:eigenstates} can be prepared through driving by a resonant laser field. Let $|g,n\rangle$ denote the ground state of the system dressed by $|n\rangle$ photons and $|e_{1/2},n-1\rangle$ the excited states with one photon less. This constitutes a $V$-level scheme. Under resonant driving, these states are all degenerate but can be coupled using the effective transition Rabi frequencies $\omega_{1/2}$ (which depend on the coupling to the field) as
$$
\hat{H}_{\rm red}=\hbar\left(
\begin{array}{lll}
0&\omega_1&\omega_2 \rme^{-\rmi\varphi}\\
\omega_1&0&0\\
\omega_2 \rme^{\rmi\varphi}&0&0\\
\end{array}
\right),
$$
where the phase $\varphi$ can be introduced by a relative phase between the driving fields. Diagonalization of this Hamiltonian leads to eigenstates that are hybridized between the coupled qubit system and the driving photons. One finds that under the corresponding time-evolution operator the application of a pulse of length $t=\pi/2\sqrt{\omega_1^2+\omega_2^2}$ to the ground state $|g,n\rangle=(1,0,0)$ leads to the final state
$$
|\psi_f\rangle=\frac{1}{\sqrt{\omega_1^2+\omega_2^2}}
\left(\begin{array}{l}
0\\
\omega_1\\
\omega_2 \rme^{\rmi\varphi}\\
\end{array}\right).
$$
Using multiple coils, arbitrary combinations of $\omega_1$ and $\omega_2$ as well as relative phases can be produced and thus, arbitrary linear combinations of the degenerate states can be prepared. This type of preparation requires multiple, phase-locked microwave drives, ideally originating from the same source. This type of phase stability is routinely achieved in high-precision microwave control \cite{Lucero08,Chow09b}. As this is essentially a precise Rabi pulse, it 
is relatively robust against decoherence. However, after the preparation of the state, 
decoherence in the degenerate subspace essentially selects the eigenstates of the coupling to the environment, projected to the subspace, which will in general be not or much more weakly entangled. Hence, although the preparation of these states is as robust as any preparation of eigenstates, their maintenance is, in general, as fragile as that of a superposition.

\section{Local Decomposition of entanglement witnesses\label{ch:appendix_local_measurements}}

Table \ref{tab_witness_local_decompositions} lists the decompositions of the used entanglement witnesses into local projective measurements. The decompositions are optimal in that they require the minimal set of local measurements (e.g. four measurement settings for $\mathcal{W}_{\overline{\rm GHZ}}^{(1)}$: $\hat\sigma_z^{\otimes 3}$ --i.e. $\hat\sigma_z$ on all qubits--, $\hat\sigma_y^{\otimes 3}$, $(\hat\sigma_z+\hat\sigma_x)^{\otimes 3}$, and $(\hat\sigma_z-\hat\sigma_x)^{\otimes 3}$).

\begin{table}[htb]
\caption{\label{tab_witness_local_decompositions}
Local decomposition of the entanglement witnesses used above. The optimal decomposition for $\mathcal{W}_{\overline{\rm GHZ}}^{(1)}$ can be found in \cite{Weinfurter2004} (four settings), the one for $\mathcal{W}_{\overline{\rm GHZ}}^{(2)}$ (four settings) was calculated from $\mathcal{W}_{\overline{\rm GHZ}}^{(2)}=\hat R_x(\pi)^{\otimes 3}\,\mathcal{W}_{\overline{\rm GHZ}}^{(1)}\, \hat R_x(-\pi)^{\otimes 3}$. The decomposition for $\mathcal{W}_{\rm \overline{W}}$ (five settings) was obtained similarly from the optimized decomposition $\mathcal{W}_{\rm W}^{(1)}$ (five settings) derived in \cite{Guehne2003}, $\mathcal{W}_{\rm \overline{W}}=\hat U^{L \, \dag}\, \mathcal{W}_{\rm W}^{(1)}\, \hat U^L$.}
\centering
\begin{tabular}{ll}
  \br
    EW & Local decomposition \\
  \mr   
    $\mathcal{W}_{\overline{\rm GHZ}}^{(1)}$ & $\frac{1}{16}\,\big[10\cdot \unity^{\otimes 3}+ 4\,\hat\sigma_z^{\otimes 3}
-2(\hat\sigma_y \, \hat\sigma_y \, \unity + \hat\sigma_y \, \unity \, \hat\sigma_y + \unity \, \hat\sigma_y \, \hat\sigma_y)-(\hat\sigma_z+\hat\sigma_x)^{\otimes 3}-(\hat\sigma_z-\hat\sigma_x)^{\otimes 3}\big]$\\[3.5ex]

    $\mathcal{W}_{\overline{\rm GHZ}}^{(2)}$ & $\frac{1}{16}\,\big[10\cdot \unity^{\otimes 3}- 4\,\hat\sigma_z^{\otimes 3}
-2(\hat\sigma_y \, \hat\sigma_y \, \unity + \hat\sigma_y \, \unity \, \hat\sigma_y + \unity \, \hat\sigma_y \, \hat\sigma_y)+(\hat\sigma_z-\hat\sigma_x)^{\otimes 3}+(\hat\sigma_z+\hat\sigma_x)^{\otimes 3}\big]$\\[3.5ex]

    $\mathcal{W}_{\rm \overline{W}}$ & $ \frac{1}{24}\,\big[ 17\cdot\unity^{\otimes 3}-7\,        \hat\sigma_x^{\otimes 3}-3(\hat \sigma_x \, \unity \, \unity + \unity \, \hat\sigma_x \, \unity + \unity \, \unity \, \hat\sigma_x)+
5(\hat\sigma_x \, \hat\sigma_x \, \unity + \hat\sigma_x \, \unity \, \hat\sigma_x + \unity \, \hat\sigma_x \, \hat\sigma_x) $\\[0.5ex]
    &  $-(\unity-\hat\sigma_x+\hat\sigma_z)\otimes(\unity-\hat\sigma_x-\frac{\sqrt{3}}{2}\,\hat\sigma_y-\frac{1}{2}\,\hat\sigma_z) \otimes(\unity-\hat\sigma_x+\frac{\sqrt{3}}{2}\,\hat\sigma_y-\frac{1}{2}\,\hat\sigma_z)-$\\[0.5ex]

    &  $-(\unity-\hat\sigma_x-\hat\sigma_z)\otimes(\unity-\hat\sigma_x+\frac{\sqrt{3}}{2}\,\hat\sigma_y+\frac{1}{2}\,\hat\sigma_z) \otimes(\unity-\hat\sigma_x-\frac{\sqrt{3}}{2}\,\hat\sigma_y+\frac{1}{2}\,\hat\sigma_z)-$\\[0.5ex]

    &  $-(\unity-\hat\sigma_x+\hat\sigma_y)\otimes(\unity-\hat\sigma_x-\frac{1}{2}\,\hat\sigma_y+\frac{\sqrt{3}}{2}\,\hat\sigma_z) \otimes(\unity-\hat\sigma_x-\frac{1}{2}\,\hat\sigma_y-\frac{\sqrt{3}}{2}\,\hat\sigma_z)-$\\[0.5ex]

    &  $-(\unity-\hat\sigma_x-\hat\sigma_y)\otimes(\unity-\hat\sigma_x+\frac{1}{2}\,\hat\sigma_y-\frac{\sqrt{3}}{2}\,\hat\sigma_z) \otimes(\unity-\hat\sigma_x+\frac{1}{2}\,\hat\sigma_y+\frac{\sqrt{3}}{2}\,\hat\sigma_z) \big]$\\[1.5ex]

  \br
\end{tabular}
\end{table} 

\bibliography{entanglement_paper_draft}

\end{document}